# Solar Thermal Energy Conversion Enhanced by Selective Metafilm Absorber under Multiple Solar Concentrations at High Temperatures

*Hassan Alshehri,[1,2] Qing Ni,[1] Sydney Taylor,[1] Ryan McBurney,[1] Hao Wang,[1,3] Liping Wang[1,*]*

[1]*School for Engineering of Matter, Transport and Energy,*
*Arizona State University, Tempe, Arizona 85287, USA*

[2]*Mechanical Engineering Department, College of Engineering,*
*King Saud University, Riyadh, Saudi Arabia*

[3]*Energy Storage and Distributed Resources Division,*
*Lawrence Berkeley National Laboratory, Berkeley, CA 94720 USA*

[*] *liping.wang@asu.edu*

## Summary

Concentrating solar power, particularly parabolic trough system with solar concentrations less than 50, requires spectrally selective solar absorbers that are thermally stable at high temperatures of 400°C above to achieve high efficiency. In this work, the solar-thermal performance of a selective multilayer metafilm absorber is characterized along with a black absorber for comparison by a lab-scale experimental setup that measures the steady-state absorber temperature under multiple solar concentrations. Heat transfer analysis is employed to elucidate different heat transfer modes and validate the solar-thermal experiment. Due to the superior spectral selectivity and excellent thermal stability, the metafilm absorber deposited on the cost-effective stainless steel foil could achieve the solar-thermal efficiency 57% at a steady-state temperature of 371°C under 10 suns during the lab-scale experiment with losses, while a highest efficiency of 83% was projected under the same conditions for practical solar thermal applications. The results here will facilitate the research and development of novel solar materials for high-efficiency solar thermal energy conversion.

# Introduction

Renewable energy technologies, such as solar energy, are becoming increasingly important and widespread in today's world as more governments and private companies adopt these technologies as an alternative to fossil fuels. Solar energy can be converted to power efficiently through two main methods: by using solar photovoltaic cells to convert the solar energy to electricity, or by using solar thermal absorbers to convert the solar energy to high-temperature heat to generate power with heat engines [1]. Solar thermal absorbers are required to be efficient at absorbing solar energy without significant thermal losses due to re-emission at high temperatures particularly in parabolic trough collectors with concentration factors less than 50 [2,3]. This can be achieved through the spectral selectivity of the absorbers, while an ideal spectrally selective solar absorber has unity absorptance in the solar spectrum and zero emittance in the infrared range where most of the thermal loss occurs [4]. Concentrating solar power (CSP) systems require solar thermal absorbers to be operated at temperatures of 400°C above to drive heat engines commonly as steam turbines for power generation [5]. Thus, spectrally selective solar absorbers that are thermally stable at high temperatures are crucial for highly efficient CSP systems.

One of the most common coatings in high-temperature CSP systems is Pyromark 2500 [6], which loses a significant amount of heat through emission in the IR range due to the lack of high spectral selectivity in spite of high solar absorption and thermal stability at elevated temperatures [7]. Artificial composites or micro/nanostructured metamaterials have been recently developed as selective solar thermal absorbers [8-14] such as subwavelength gratings [15-20], nanocomposites [21] or nanoparticles [22], cermet [23-25], photonic crystals [26-27], and multilayers [28-31]. Wang *et al*. recently reported the design and fabrication of an ultrathin multilayer selective solar absorber [30], namely metafilm absorber, which is thermally stable in air up to 600°C, while thermal cycle testing revealed its long-term thermal stability at 400°C in ambient conditions.

The basic working principle of solar thermal collectors is illustrated in **Figure 1A**. Concentrated (or unconcentrated for flat-plate collectors) sunlight is incident onto the solar absorber at the collector surface, which results in a significant temperature rise. The absorber then conducts heat to a constantly flowing heat transfer fluid like water, oil, or molten salt depending on the temperature ranges for particular applications. The stagnation temperature of a solar thermal collector, which is the steady-state temperature when the heat transfer fluid is not flowing (or stagnant), indicates the maximum temperature an absorber can achieve as a measure for solar thermal performance. According to Duffie and Beckman [32], the stagnation temperature of a single cover flat-plate (i.e., 1 sun) collector ranges from 150°C for nonselective coatings to 300°C for highly selective coatings. Standard EN12975, developed for solar thermal collector testing, includes guidelines for measuring the stagnation temperature of collectors, including flat-plate collectors and evacuated tube collectors under solar irradiance of 1000 W/m$^2$ [33]. Previous works [34-36] were focused on measuring the stagnation temperature of full-sized solar collectors mainly with natural coatings. There are few studies on the characterization of novel metamaterial selective absorbers to experimentally evaluate their solar thermal performance under different solar concentrations due to the challenges associated with smaller absorber sample sizes developed in labs mainly oriented towards optical characterizations and thermal tests. A lab-scale solar thermal test is in urgent need to experimentally confirm enhanced solar thermal performance with small-sized metamaterial absorbers before mass manufacturing and outdoor field testing on full-scale solar collectors.

## Results

In this work, the solar-to-heat performance of previously-deveopled selective metafilm absorbers, deposited on both a polished silicon wafer and a cost-effective and flexible stainless-steel foil, was characterized experimentally via a lab-scale setup that measures the steady-state absorber temperature under multiple solar concentrations. A black absorber as a reference sample was also measured for comparison. The optical and radiative properties of the absorbers were measured after multiple thermal cycling tests at high temperatures up to 700°C in vacuum. A heat transfer model was used to quantify the solar-thermal efficiency at the measured steady-state temperature of the absorbers by considering the conductive heat loss as useful heat gain, for which experimental and theoretical values are both presented. A simple cost analysis was also performed to compare the material cost of the metafilm absorber with some commercial solar coatings.

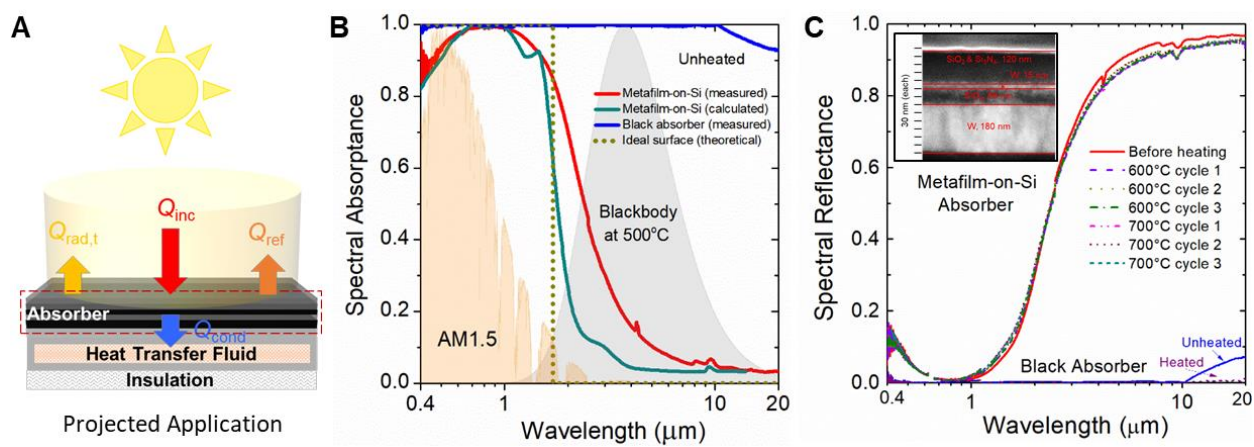

**Figure 1 A**. Illustration of projected solar-thermal application with heat transfer modes for the solar absorber. **B.** Spectral absorptance of the metafilm-on-Si absorber (measured and theoretical calculation), a black absorber (measured), and an ideal absorber (theoretical) before heating along with AM1.5 solar irradiance and blackbody spectra at 500°C. **C.** Spectral reflectance of the metafilm-on-Si absorber and the black absorber before and after multiple thermal cycling tests at 600°C and 700°C. Inset shows the cross-sectional SEM image of the metafilm structure.

**Figure 1B** shows the normal spectral absorptance for the fabricated metafilm-on-silicon sample over a wavelength range from 0.4 μm to 20 μm from theoretical calculations and spectroscopic reflectance measurements (see details in **Experimental Procedures** and **Figure S1, S2** for spectral hemispherical and diffuse reflectance). The selective metafilm absorber consisting of a Fabry-Perot cavity (W-SiO$_2$-W) with two anti-reflective coatings on top (Si$_3$N$_4$ and SiO$_2$) was fabricated on a silicon wafer following the processes reported in an earlier work [30] (see details in **Experimental Procedures**). The inset of **Figure 1C** presents the cross-sectional scanning electron microscope (SEM) image of the fabricated metafilm-on-Si sample with the material and thickness labeled for each layer. Note that the selective metafilm absorber exhibits excellent spectral selectivity compared with an ideal selective solar absorber that has unity absorptance in the solar spectrum and zero emittance in the infrared region with an optimum cutoff of 1.7 μm at 500°C. After spectral integration, the metafilm-on-Si absorber has very high total solar absorptance of $\alpha = 0.942$ within the solar irradiance spectrum (AM1.5) [37], and low total emittance of $\varepsilon = 0.153$ within the blackbody emission spectrum at 500°C (see **Figure S3** for temperature-dependent total emittance). The measured spectral absorptance of unheated and heated black absorbers (Acktar Metal Velvet) is also presented for comparison as some commercial solar coatings like Pyromark 2500 have similar broadband near-zero reflectance. From **Figure 1B**, it can be seen that the black absorber has almost 100% absorptance throughout the visible, near-infrared, and mid-infrared ranges with a slight decrease beyond the 10 μm wavelength, as reported by the manufacturer [38], resulting in a total solar absorptance of 0.998 and a total emittance of 0.936 at 500°C. Due to its significantly lower emittance compared to the black absorber, the metafilm absorber would possibly achieve higher solar-thermal conversion efficiency with much smaller thermal emission loss.

Thermal cycling tests at 600°C and 700°C were conducted for the metafilm-on-Si absorber, which went through 3 heating/cooling cycles with each consisting of 5 hours of heating in vacuum and 5 hours of cooling to room temperature (see details in **Experimental Procedures** and **Figure S4** for the experimental setup of the thermal cycling test). As shown in **Figure 1C**, the spectral reflectance, which was measured after each cycle at room temperature with an FTIR spectrometer, shows little change after 3 cycles at both 600°C and 700°C. This confirms the excellent thermal stability of the fabricated metafilm absorbers at high temperatures. Heating tests at higher temperatures were not performed due to the limitation of the heater that was used.

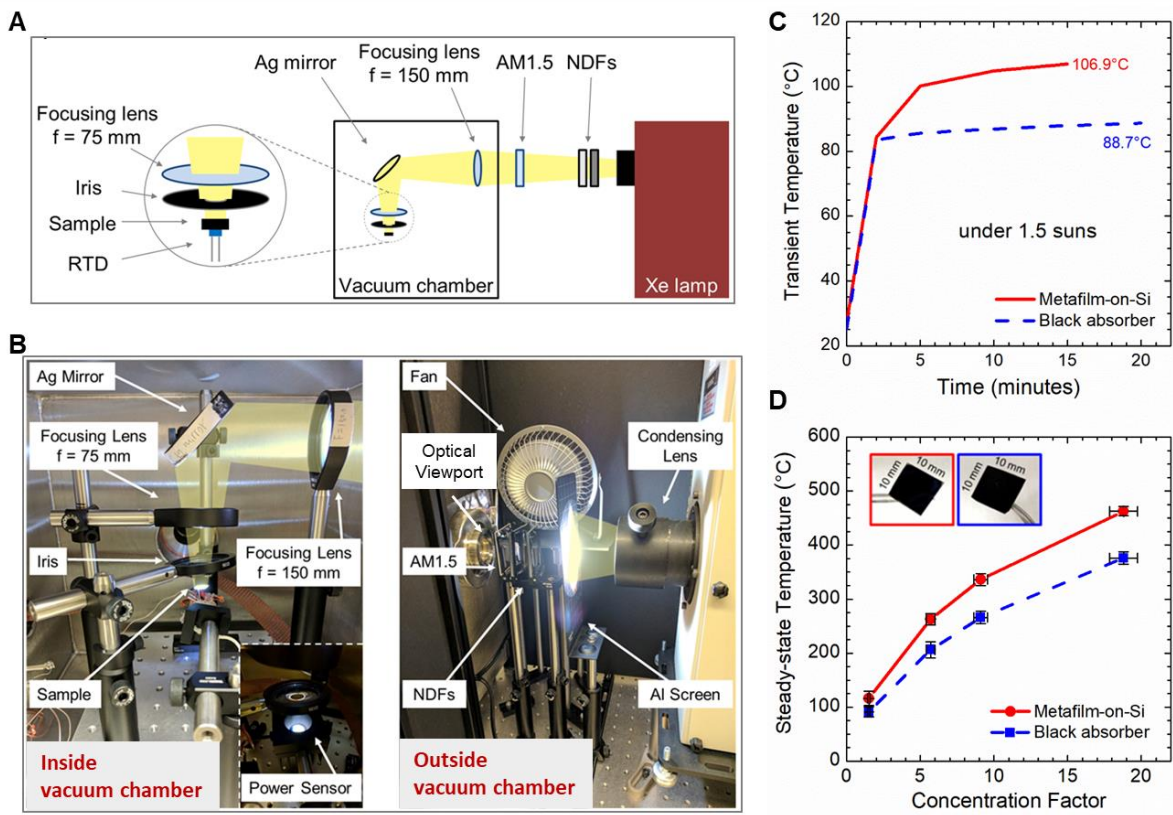

**Figure 2 A.** Schematic of the solar-thermal experiment setup. **B.** Photos of major optical and mechanical parts of the experimental setup. **C.** Transient sample temperatures for the solar thermal experiments of the metafilm and black absorbers under 1.5 suns. **D.** Steady-state temperatures under multiple suns. The tested absorber samples are shown in the inset.

A solar-thermal experiment setup was built to experimentally characterize the solar-thermal conversion with these selective metafilm absorbers. A schematic of the setup is shown in **Figure 2A** and photos of major optical and mechanical parts are shown in **Figure 2B**. The setup can be used to test solar absorbers with a size of 10 mm × 10 mm or so under different solar concentrations from 1.5 suns up to 20 suns using a xenon arc lamp, an AM1.5 filter, and multiple neutral density filters. A resistance temperature detector (RTD) was used to measure the absorber temperature by attaching it to the backside of the sample with thermal paste. The experiments were conducted under a vacuum pressure less than $3\times10^{-3}$ Torr to minimize thermal convection losses (see details in **Experimental Procedures**).

**Figure 2C** shows the transient temperature profile from one solar-thermal experiment under 1.5 suns for the metafilm-on-Si absorber and the black absorber, where the metafilm reaches a higher steady-state temperature (106.9°C) than the black absorber (88.7°C). **Figure 2D** presents the steady-state temperatures under different solar concentrations (i.e., 1.5, 5.7, 9.1, and 18.8 suns) with each averaged from three independent measurements along with error bars (see details on uncertainty analysis in the Supplemental Information). It can be seen that the metafilm outperforms the black absorber at every concentration factor due to its much lower emittance in the infrared region. This is more significant at high concentrations where higher temperatures lead to more severe radiation losses from the black absorber, resulting in a larger difference in its steady-state temperatures compared to the metafilm absorber. In particular, the metafilm reached a steady-state temperature of 463°C at 18.8 suns, which is 87°C higher than that achieved by the black absorber at the same concentration factor. This strongly indicates enhanced solar-thermal conversion performance with the low-emitting selective metafilm absorbers.

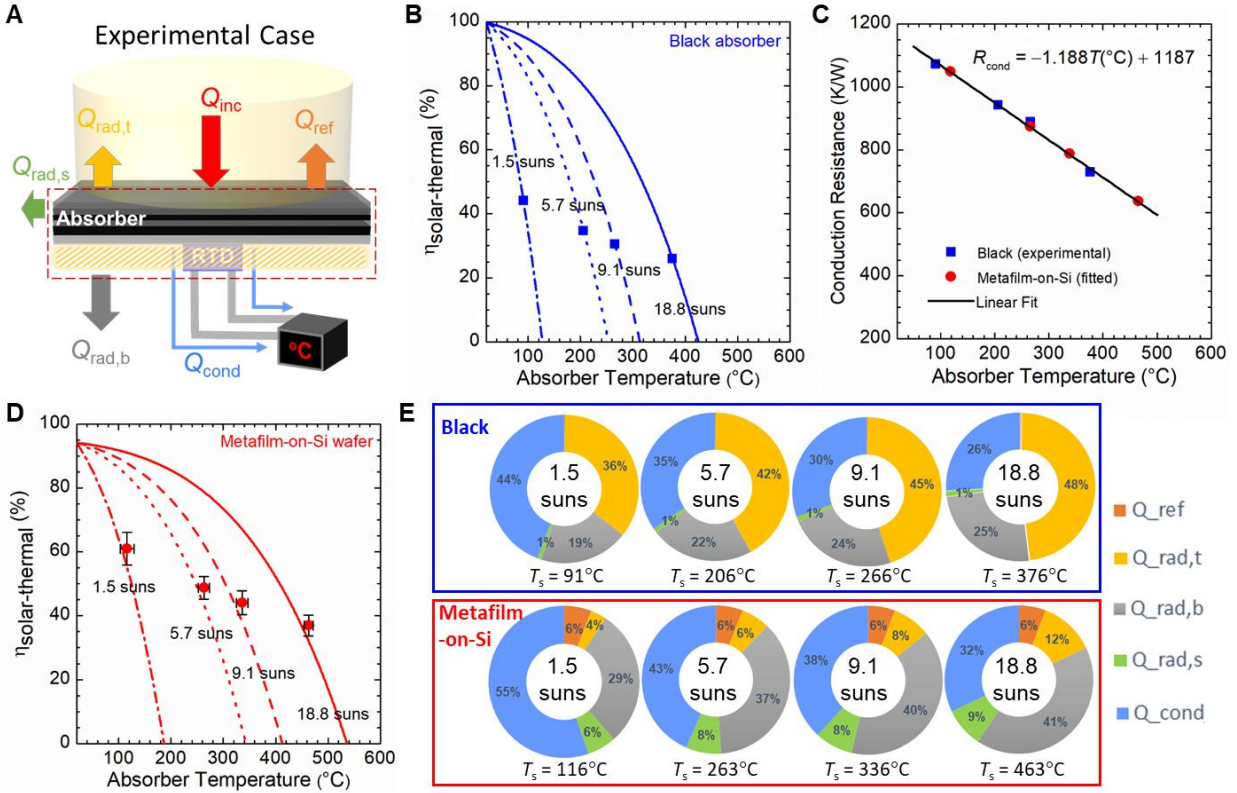

**Figure 3 A.** Illustration of heat transfer modes during the solar-thermal experiment. **B**. Solar-thermal conversion efficiencies of black absorber under 1.5, 5.7, 9.1, and 18.8 suns with markers from experiment and lines from theoretical calcualtion. **C.** Linearly-fitted RTD conduction resistance as a function of absorber temperature based on the solar-thermal measurement of the black absorber. **D.** Experimental (markers with error bars) and theoretical solar-thermal efficiency of metfilm-on-Si absorber under 1.5, 5.7, 9.1, and 18.8 suns. **E.** Energy analysis pie charts of heat transfer modes during solar-thermal measurements for black absorber (blue solid box) and metafilm-on-Si absorber (red solid box) under multiple suns along with resulting steady-state absorber temperature.

To quantitatively understand the solar-thermal conversion, different heat transfer modes for the absorber under steady-state conditions during the solar-thermal experiment were analyzed as illustrated in **Figure 3A**. Incident solar energy after concentration is $Q_{\text{inc}} = A_t C G$, where $A_t$ is the top surface area of the absorber, $C$ is the concentration factor, and $G$ is the solar constant (1000

W/m$^2$) from AM1.5. Reflected energy by the top absorber surface is $Q_{\text{ref}} = RQ_{\text{inc}}$, where $R$ is the total reflectance of the metafilm or black absorber and can be calculated by subtracting the total solar absorptance from unity ($R = 1-\alpha$). Radiation losses from the absorber to the environment can be calculated as $Q_{\text{rad},i} = A_i \varepsilon_i \sigma (T_s^4 - T_\infty^4)$, where $i$ indicates the top (t), side (s), and bottom (b) surfaces of the absorber, $A_i$ and $\varepsilon_i$ are the area and (temperature-dependent) total emissivity of the corresponding surfaces, and $T_\infty = 20°C$ is the temperature of the vacuum chamber wall. Note that the top surface is either a metafilm or a black absorber with the respective total emittance, while the side and bottom surfaces of both absorbers were fully covered by thermal paste (see **Figure S1-S3** for the measured spectral reflectance and the calculated total emittance at different temperatures for the metafilm, black absorber, and thermal paste). Finally, the heat conducted through the RTD wires, which is potentially useful heat gain converted from solar energy and transported by heat transfer fluids in practice, is considered as $Q_{\text{cond}} = (T_s - T_\infty)/R_{\text{cond}}(T_s)$, where $R_{\text{cond}}(T_s)$ is the conduction resistance that is dependent on the absorber temperature $T_s$. Here, the black absorber was first measured as a reference to find the correlation between $R_{\text{cond}}(T_s)$ and $T_s$, and then the conduction heat gain by the metafilm absorber can be calculated from the correlation based on the measured steady-state temperature at a given solar concentration. All the parameters, such as the areas and emittance values at different temperatures, used in the solar-thermal analysis of the metafilm-on-Si absorber were tabulated in **Table S1**.

The theoretical solar-thermal conversion efficiency $\eta_{\text{theo}}$ can be calculated based on the theoretical heat gain from the energy balance in the heat transfer model as

$$\eta_{\text{theo}} = \frac{Q_{\text{gain,theo}}}{Q_{\text{inc}}} = \frac{Q_{\text{inc}} - Q_{\text{ref}} - \Sigma_{t,b,s} Q_{\text{rad}}}{Q_{\text{inc}}} \quad (1)$$

while the solar-thermal efficiency from the experiment can be found based on the conducted heat via the RTD wires as

$$\eta_{\exp} = \frac{Q_{\text{gain,exp}}}{Q_{\text{inc}}} = \frac{Q_{\text{cond}}}{Q_{\text{inc}}} = \frac{T_s - T_\infty}{R_{\text{cond}}(T_s) Q_{\text{inc}}} \tag{2}$$

Note that in the experiment, there is no direct way of measuring heat conduction or conduction resistance from the RTD wires. Therefore, the black absorber is used as a calibration to find $R_{\text{cond}}(T_s)$ by letting $\eta_{\text{theo}} = \eta_{\exp}$ at the same steady-state temperature $T_s$ under a given solar concentration. As shown in **Figure 3B**, the black absorber achieves solar-thermal efficiencies of 44% at 91°C under 1.5 suns, 35% at 206°C under 5.7 suns, 30% at 266°C under 9.1 suns, and 26% at 376°C under 18.8 suns. Then a linear fit between the $R_{\text{cond}}$ and the absorber temperature $T_s$ was obtained as $R_{\text{cond}} = -1.188 T_s + 1187$ based on the black absorber as shown in **Figure 3C**. Further, the linear fit is used to calculate $R_{\text{cond}}$ for the metafilm absorber at its measured steady-state temperatures, from which the experimental solar-thermal efficiency $\eta_{\exp}$ for the metafilm absorber was obtained under different solar concentrations from Eq. (2).

As presented in **Figure 3D**, the experimental solar-thermal efficiencies (markers with error bars) of the metafilm-on-Si absorber are in excellent agreement with the theoretical ones calculated from Eq. (1). In particular, the metafilm-on-Si absorber achieves experimental solar-thermal efficiencies of 61% at 116°C under 1.5 suns, 49% at 263°C under 5.7 suns, 44% at 336°C under 9.1 suns, and 37% at 463°C under 18.8 suns. In comparison with the black absorber, the metafilm-on-Si absorber achieves 11% more solar-thermal efficiency at an 87°C higher absorber temperature under the same 18.8 suns solar concentration during the solar-thermal experiment.

To better understand the solar-thermal performance during the experiment, **Figure 3E** shows the pie charts of all the radiation and conduction heat transfer normalized to the incident solar energy in percentage calculated for both the black absorber and the metafilm-on-Si absorber based on the heat transfer model. For the black absorber, most of the heat loss is from thermal radiation at the top surface (i.e., 36% to 48% from 1.5 to 18.8 suns) due to its almost unity emittance. On the other hand, for the metafilm the radiation loss from the top surface is minimal (i.e., 4% to 12%) due to the low emittance archived by its excellent spectral selectivity. However, most of the heat loss for the metafilm absorber during the solar-thermal experiment was via radiation from the bottom surface (i.e., 29% to 41%) and from the sides (i.e., 6% to 9%), both of which were covered by thermal paste with a high emittance of around 0.5.

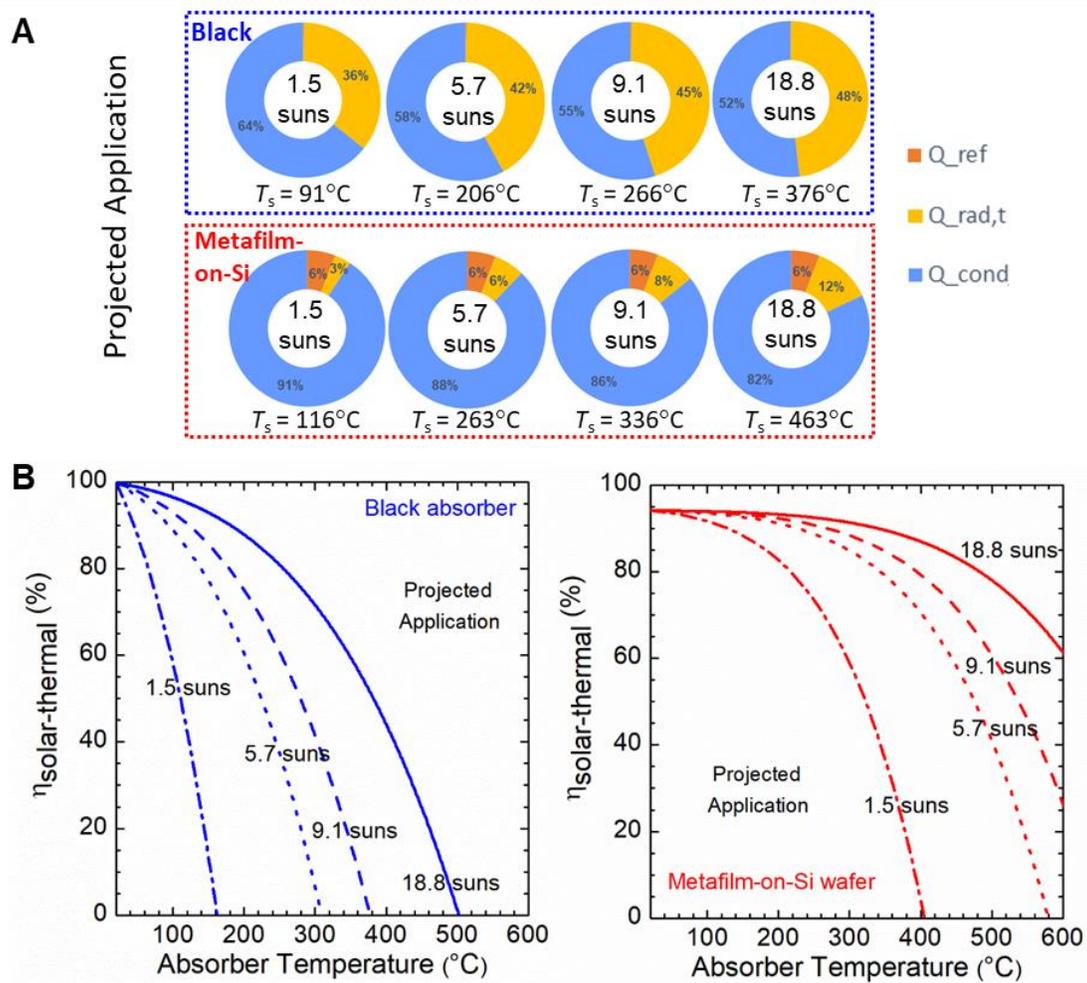

**Figure 4 A.** Energy analysis pie charts of the projected solar-thermal application for the black absorber (blue dashed box) and the metafilm-on-Si absorber (red dashed box) under 1.5, 5.7, 9.1, and 18.8 suns along with the corresponding steady-state absorber temperatures. **B.** Theoretical solar-thermal efficiency of the black absorber (blue curves) and the metafilm-on-Si absorber (red curves) for the projected solar-thermal application under multiple suns.

As a matter of fact, the radiation losses from the bottom or sides of the solar absorber coatings could potentially be eliminated by conformally covering thermally insulated tubes filled with a high-temperature heat transfer fluid in practical solar-thermal applications as illustrated in **Figure 1A**. In such a projected application, the solar-thermal efficiency can be defined as

$$\eta_{\text{proj}} = \frac{Q_{\text{gain,proj}}}{Q_{\text{inc}}} = \frac{Q_{\text{inc}} - Q_{\text{ref}} - Q_{\text{rad,t}}}{Q_{\text{inc}}} \tag{3}$$

where $Q_{\text{ref}}$ and $Q_{\text{rad,t}}$ are the only heat loss modes. The pie charts in **Figure 4A** show the heat gain, radiation loss, and reflected energy in percentage for both the black absorber and the metafilm-on-Si absorber for the projected application at the same steady-state temperatures and solar concentrations from the solar-thermal experiment. The black absorber is predicted to achieve projected solar-thermal efficiency values between 64% and 52% from 1.5 to 18.8 suns with an increase from 20% to 26% compared to the solar-thermal experiment, while the radiation loss from the top surface remains the same (from 36% to 48%). On the other hand, the metafilm-on-Si absorber could achieve a theoretical solar-thermal efficiency between 91% and 82% from 1.5 to 18.8 suns for the projected application with an increase of 35% up to 50% compared to the solar-thermal experiment after the bottom and side radiation losses are eliminated. **Figure 4B** plots the theoretical solar-thermal efficiency at different absorber temperatures for the projected application for both the black and metafilm-on-Si absorbers under 1.5, 5.7, 9.1, and 18.8 suns. Clearly, the metafilm absorber outperforms the black one with a much higher solar-thermal efficiency under the same temperature (higher than 150°C) and solar concentration. In particular, under the same 18.8 suns, the black absorber reaches a stagnation temperature of 500°C with zero solar-thermal efficiency, while the metafilm-on-Si absorber converts 78% of the incident solar energy to useful heat at 500°C, achieving a solar-thermal efficiency of 61% at 600°C.

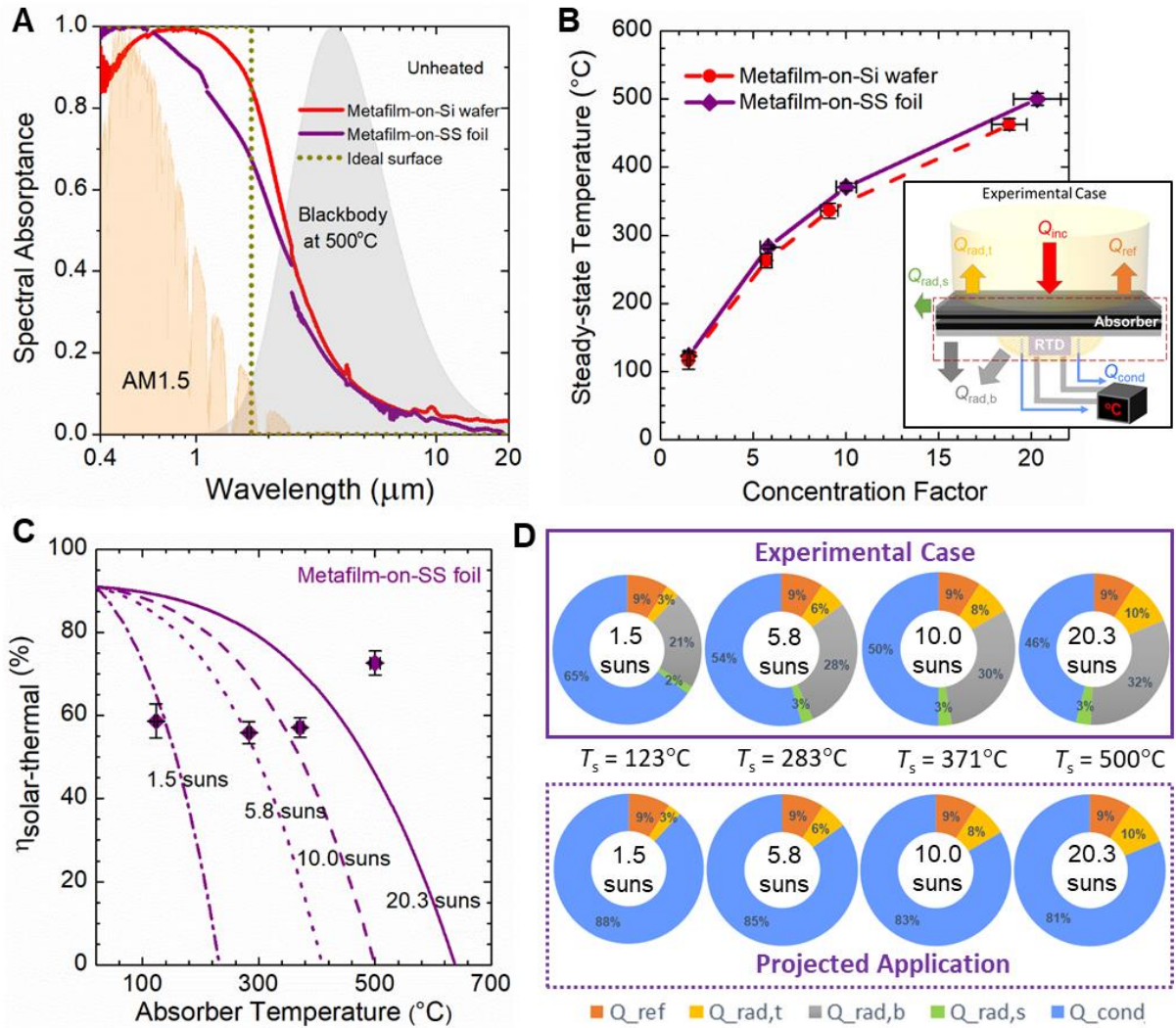

**Figure 5 A.** Measured spectral absorptance of metafilms deposited on stainless-steel foil (metafilm-on-SS foil) and on silicon wafer (metafilm-on-Si wafer) before heating. **B.** Steady-state temperatures of metafilm-on-SS-foil and metafilm-on-Si-wafer absorbers under different concentration factors from the lab measurement with uncertainty. Inset illustrates the heat transfer model for the metafilm-on-SS-foil sample during the solar-thermal experiment. **C.** Solar-thermal conversion efficiencies of the metafilm-on-SS-foil absorber under 1.5, 5.8, 10.0, and 20.3 suns with markers from the experiment and lines from the theoretical calculation. **D.** Energy analysis pie charts of the metafilm-on-SS-foil absorber for the solar-thermal lab experiment (purple solid box) and projected application (purple dashed box) under multiple suns.

To further demonstrate the potential of the metafilm absorber in practical applications, the metafilm structure was deposited onto a polished flexible stainless-steel foil (metafilm-on-SS). **Figure 5A** shows the measured spectral absorptance of an unheated metafilm-on-SS sample, where excellent spectral selectivity is again demonstrated with a small difference in the absorptance spectrum compared to the metafilm-on-Si absorber. As a result, the metafilm-on-SS sample has a slightly lower solar absorptance of $\alpha = 0.910$ and a smaller total emittance $\varepsilon = 0.100$ at 500°C than the metafilm on the Si wafer. Steady-state temperatures under 1.5, 5.8, 10.0, and 20.3 suns from the solar-thermal experiments for the metafilm-on-SS absorber are presented in **Figure 5B,** which are notably higher than the metafilm-on-Si absorber. As depicted in the inset, the thermal paste on the backside of the metafilm-on-SS foil sample only covered the RTD sensor, while about half of the bottom surface of the stainless-steel foil was exposed. The stainless-steel foil has a much smaller emittance (~0.12) compared to the thermal paste (~0.5). Again, a black absorber was used for fitting the $R_{\text{cond}}$ as a function of $T_s$ for the metafilm-on-SS sample during the solar-thermal experiments. **Figure S5** presents the solar-thermal efficiency during the experiment and for the projected application along with energy analysis pie charts for the black absorber, while **Figure S6A** shows the linear fit of $R_{\text{cond}}(T_s)$ based on the black absorber and calculated $R_{\text{cond}}$ for the metafilm-on-SS absorber. All the parameters, such as the areas and total emittances at different temperatures, used in the solar-thermal analysis of the metafilm-on-SS absorber were tabulated in **Table S2**.

**Figure 5C** shows the measured solar-thermal efficiencies with uncertainty (i.e., markers with error bars) of the metafilm-on-SS absorber, which are in excellent agreement with the theoretical prediction except for 20.3 suns. In particular, the metafilm-on-SS absorber achieves experimental solar-thermal efficiencies of 59% at 123°C under 1.5 suns, 56% at 283°C under 5.8 suns, and 57% at 371°C under 10.0 suns. The metafilm-on-SS absorber achieved a steady-state temperature of about 20°C or 35°C higher along with a 7% or 13% higher efficiency under 5.7 suns or 10.0 suns compared to the metafilm-on-Si absorber under 5.8 suns or 9.1 suns. At 20.3 suns, the metafilm-on-SS absorber reached an experimental solar-thermal efficiency of 73% at 500°C, which was the temperature limit of the RTD sensor used. The large discrepancy from the theory (27% higher than the prediction) is believed to be due to the melting of the RTD packaging materials whose latent heat increases the solar-thermal efficiency during the experiment.

The higher solar-thermal performance of the metafilm-on-SS absorber compared to the metafilm-on-Si absorber from the lab-scale experiments can be understood from the theoretical energy analysis presented as pie charts in **Figure 5D**. For the experimental case, although the metafilm-on-SS absorber has a slightly higher percentage in solar reflection (i.e., 9%), the radiation loss from the bottom surface takes 21% up to 32% if the incident solar energy, which is about 8% or 9% less than the metafilm-on-Si absorber. Moreover, the percentage of the side radiation loss from the metafilm-on-SS absorber is only 2% or 3%, while the metafilm-on-Si absorber has a side loss of 8% or 9% of the incident solar energy. This is because less high-emitting thermal paste was used to only cover the RTD sensor while the low-emitting stainless-steel had about half of the bottom surface exposed. For the projected application without bottom and side radiation losses, the metafilm-on-SS absorber is predicted to achieve 88% at 123°C under 1.5 suns, 85% at 283°C under 5.8 suns, 83% at 371°C under 10.0 suns, and 81% at 500°C under 20.3 suns (see **Figure S6B** for predicted solar-thermal efficiency as a function of absorber temperature under different solar concentrations for the projected application).

Finally, a cost analysis was conducted to compare the metafilm to other concentrating solar coating materials. The developed metafilms were fabricated at university facilities rather than commercial-scale setups. Therefore, a reasonable comparison of equipment costs as well as labor costs with commercial solar coatings cannot be made. Instead, a material cost analysis of the metafilm was performed (see details in the Supplemental Information), which turned out to be $2.23 per square meter. This is significantly lower than the materials cost of some commercial solar absorber coatings such as Pyromark, $Co_3O_4$, and LSM which cost $5.41, $50, and $100 per square meter, respectively [39].

## Discussion

The solar-thermal energy conversion performance of selective metafilm absorbers with excellent spectral selectivity and high-temperature thermal stability was successfully characterized with a lab-scale experimental setup under multiple solar concentrations up to 20 suns at temperatures up to 500°C. The solar-thermal efficiency of a flexible cost-effective metafilm-on-stainless-steel-foil absorber was experimentally characterized to be 57% at 371°C under 10 suns from a lab-scale setup. The efficiency was projected to reach 83% in practical solar thermal applications by eliminating side and bottom radiation losses. Higher solar concentrations and high temperatures could possibly be reached with a more powerful solar simulator and more durable temperature sensors. The developed experimental setup and approach, along with the theoretical analysis, enables lab-scale performance characterization of novel solar-thermal materials under multiple solar concentrations before large-scale manufacturing and field tests. This would facilitate the research and development of high-efficiency solar-thermal energy harvesting and conversion, in particular CSP, with reduced costs.

**Experimental Procedures**

*Sample Fabrication.* The bottom three layers of the metafilm (W-SiO$_2$-W) were deposited by sputtering and the top two layers (Si$_3$N$_4$ and SiO$_2$) were deposited by chemical vapor deposition following the fabrication processes discussed in Wang et al. [30].

*Optical and Radiative Property Characterizations.* Spectral specular reflectance was characterized by a Fourier Transform Infrared (FTIR) spectrometer (Thermo Fisher Scientific, Nicolet iS50) with a variable-angle reflectance accessory (Harrick Scientific Products Inc., Seagull) at near-normal direction without polarizers. Spectral hemispherical ($R'^{\frown}_\lambda$) and diffuse reflectance measurements at normal incidence for the metafilm and the black absorbers (Acktar, Metal Velvet) were conducted in the wavelengths from 350 nm to 1.8 μm using a tunable light source (Newport, TLS-250QU) and an integrating sphere made of PTFE as well as in the mid-infrared with a diffuse-gold integrating sphere. The spectral emittance and absorptance were calculated based on the measured spectral reflectance as $\varepsilon'_\lambda = \alpha'_\lambda = 1 - R'^{\frown}_\lambda$ according to Kirchhoff's law and energy balance.

*Thermal Cycling Tests.* A thermal cycling testing apparatus was built to test the durability of metafilm absorbers under high temperatures in vacuum. **Figure S4A** illustrates the setup with a schematic and **Figure S4B** shows a photo of it. A Si$_3$N$_4$ heater (Induceramic) was used to heat a copper chip, whose temperature was measured by inserting a K-type thermocouple from a side hole. Thermally conductive paste (Aremco, Pyro-Duct 597-A) was used between the heater and the copper chip, to minimize contact resistance. The metafilm-on-Si sample was placed on the copper chip, while a glass slide was used to cover and protect the top surface of the metafilm from possible contamination due to outgassing inside the vacuum chamber.

*Solar Thermal Measurements.* A 1 kW xenon arc lamp (Newport 6271 in a Newport 66921 housing), which along with an AM1.5 filter (Newport, 81094) was used as a solar simulator, was operated steadily with a power supply (Newport, OPS-A1000) at a set power of 907 W. Light emitted from the lamp goes through selected neutral density filters with different transmittance values from 80% down to 10% to achieve different concentration factors from 1.5 to 20 suns or so. A fan was used to cool down the filters and protect the filter coatings from deterioration due to excessive heating. The simulated sunlight at different concentrations went into a box vacuum chamber (Kurt J. Lesker, BX1212S) through a sapphire window, and was focused onto the solar absorber at a spot size of 1 cm or so controlled by an adjustable iris (Thorlabs, ID36). A resistance temperature detector (RTD, Omega, F2020-100-B-100) at a size of 2 mm was attached onto the bottom surface of the absorber sample approximately in 10 mm × 10 mm with thermally conductive paste (Aremco, Pyro-Duct 597-A). The RTD was then connected to a temperature monitor (MYPIN, TA4) to measure the absorber temperature. A power sensor (Thorlabs, S310C) along with a power meter (Thorlabs, PM100D) was used to measure the power density of the simulated sunlight incident on the absorber. Moreover, the ambient temperature inside the vacuum chamber was monitored by a K-type thermocouple (Omega, 5TC-TT-K-30-36) placed on the inner wall of the chamber. The steady-state absorber temperatures were measured under a vacuum pressure less than $3\times10^{-3}$ Torr achieved by a turbo vacuum pump (Agilent, TPS-compact). The experiment was independently repeated three times for each of the metafilm or black absorbers.

## Acknowledgments

This work was mainly supported by National Science Foundation (NSF) under Grant No. CBET-1454698 (H.W. and L.W.), Air Force Office of Scientific Research with Grant No. FA9550-17-1-0080 (Q.N. and L.W.), and in part by NASA Space Technology Research Fellowship #NNX16AM63H (S.T.). H.A. would like to thank King Saud University and the Saudi Arabian Cultural Mission (SACM) for their sponsorship for his Ph.D. study at ASU. Access to the NanoFab and Eyring Materials Center at Arizona State University (ASU) for sample fabrication and materials characterizations was supported in part by NSF contract ECCS-1542160. R.M. is grateful to the support by the ASU Fulton Undergraduate Research Initiative.

**Author Contributions**

H.A., Q.N., H.W., and L.W. developed the solar-thermal experimental setup; H.A. and R.M. conducted the solar-thermal experiment; H.A., Q.N., and L.W. performed heat transfer analysis; H.A. and H.W. measured the optical and radiative properties; S.T. fabricated the metafilm samples; H.A. did the thermal cycling tests; H.A. and L.W. wrote the manuscript; L.W. supervised the entire project; all the authors reviewed and approved the manuscript.

**Declaration of Interests**

The authors declare no competing interests.